\documentclass[%
preprint,
showpacs,
showkeys,
preprintnumbers,
nofootinbib,
 amsmath,amssymb,
]{revtex4-1}
\usepackage[english]{babel}
\usepackage{amsfonts}
\usepackage{amsmath}
\usepackage{wasysym}
\usepackage{amssymb}
\usepackage{booktabs}
\usepackage{graphicx}
\usepackage{multirow}
\usepackage[papersize={25cm,35cm}]{geometry}
\usepackage[colorlinks=true,linkcolor=blue,citecolor=green,menucolor=blue]{hyperref}
\usepackage{url}
\usepackage[latin1]{inputenc}

\begin{document}
\preprint{cond-mat.mtrl-sci}
\title[Calibration of oscillation amplitude in Dynamic Scanning Force Microscopy]{Calibration of oscillation amplitude in Dynamic Scanning Force Microscopy}
\author{Juan Francisco Gonz\'{a}lez Mart\'{i}nez}
\email{jfgm@um.es}
\author{In\'{e}s Nieto-Carvajal}
\email{inc2@um.es}
\author{Jaime Colchero Paetz}
\email{colchero@um.es}
\affiliation{}
\date{\today}

\begin{abstract}

A method to precisely calibrate the oscillation amplitude in Dynamic Scanning
Force Microscopy (DSFM) is described. It is experimentally shown that a typical
electronics used to process the dynamic motion of the cantilever can be
adjusted to transfer the thermal noise of the cantilever motion from its
resonance frequency to a much lower frequency within the typical bandwidth of
the corresponding electronics. Therefore, the thermal noise measured in the
in--phase (``phase'') and out--of--phase (``amplitude'') output of such an
electronics can be related to the thermal energy $kT$. If the force constant of
the cantilever is known then the oscillation amplitude can be precisely
calibrated from the thermal power measured in these signals. Based on this
concept, two procedures for the calibration of the oscillation amplitude are
proposed. One is based on simple calculation of the Root Mean Square (RMS)
measured at the outputs of the electronics used to process the dynamic motion
of the cantilever, and the other one is based on analysis of the corresponding
spectrum and the calculation of the quality factor, the resonance frequency and
the signal strength.
\end{abstract}

\maketitle

\section{Introduction}

Scanning Force Microscopy (SFM) is an extremely versatile tool for Nanoscience
that has a very high resolution, but usually only a modest precision. When SFM
is used as a microscope, precise calibration of piezo movements is necessary,
as well as careful control of piezo hysteresis and non--linearities. When used
as a force measuring instrument a precise calibration of the displacement
detector as well as the force constant of the cantilever is required. Dynamic
SFM (DSFM) is by now probably the most extended SFM mode, since it allows
operation at very low forces, and operates in the non--contact regime
\cite{Albrecht,Durig,Giessibl,GarciaR}. In DSFM the frequency (or phase) of the
tip--sample system and the oscillation amplitude are the basic signals that are
measured and should thus be calibrated for precise measurements. Precise
measurement of frequency is easy and is implemented directly in most
instruments. Calibration of oscillation amplitude is an issue that has received
little attention, and less than what we believe it deserves. In Frequency
Modulation DSFM (FM--DSFM) the frequency shift is related to (conservative)
tip--sample interaction and the oscillation amplitude to dissipation. For
Amplitude Modulation (AM--DSFM or ``tapping mode'') this separation is less
evident, nevertheless the oscillation amplitude is an important parameter,
since it critically determines non--linearity of tip--sample interaction \cite{Aime1,DynNonLinear}, onset
of bi--stability and chaos \cite{Raman1,Stark1,ChaosAFM} as well as the precise interaction  regime
(attractive vs. repulsive) used for imaging \cite{Paulo}. Finally, as shown
recently, if correctly calibrated the oscillation amplitude can be used to
obtain a reliable image of the ``true topography'' that is independent of
feedback parameters and scanning speed \cite{JuanFranPi}. In the present work
we will discuss a method to precisely calibrate the oscillation amplitude in
DSFM. Essentially, the method relates the thermal noise measured in the
``amplitude'' and ``phase'' outputs of the electronics to the thermal energy
$kT$. If the force constant of the cantilever is known then the oscillation
amplitude can be calibrated from the thermal power measured in these signals.

\section{Calibration procedures}

In the present work we discuss the response of the DSFM Dynamic Unit (DSFM--DU) in the frequency
domain because we will analyze the response of the system to thermal noise
$z_{th}(t)$. Unfortunately, thermal noise is non--trivial to describe using a
``coherent'' superposition of harmonic signals, as it is implicitly assumed in the Fourier Transform
$z(t)=\int d\nu \; z(\nu )e^{i2\pi \nu t}$. The correct description of noise in the time
domain should involve wavelet transforms \cite{wavelet}, which is however beyond of the
scope of the present work. We note that in the context of the
present work the DSFM--DU is used in a non--standard way, where no
oscillation is applied to excite the cantilever, and the reference
oscillator frequency $\nu_{r}$ is not at the resonant frequency of the
cantilever. Instead, the DSFM--DU is adjusted so that $\nu_{r}\approx \nu_{0}-bw/2$
where $bw=1/\tau$ ($\tau$: time constant) is the bandwidth of
the DSFM--DU. As discussed in detail in Appendix \ref{AppendixA}. The thermal peak of the cantilever appears centered
within the spectral range of the output of the DSFM--DU.

Essentially, the reason for this frequency shift is that when the normal force signal is internally
multiplied with the reference frequency $\nu_r$ of the DSFM--DU, frequency components at
$\nu_0-\nu_r$ and $\nu_0+\nu_r$ appear (for example from terms like $\cos (2\pi\nu_0 t)\cos (2\pi\nu_r t)$,
see also figure \ref{Fig1} for a graphical representation). Usually, the DSFM--DU is
opperated with $\nu_0=\nu_r$, then thermal noise is observed at very low frequencies $\nu\approx 0$, that is, at DC. At this low frequency
the thermal noise is usually not clearly distinguised from  pink noise ($1/\nu$--noise) of typical electronic components. However, when $\nu_0\neq\nu_r$, the thermal peak is shifted to the difference frequency $\nu_0-\nu_r$, which may be well below the resonance frequency, $\nu_0$. As discussed above, we typically adjust $\nu_r$ so that $\nu_0-\nu_r\approx bw/2\approx 4\ \rm{kHz}$ in our system.

Figure \ref{Fig2} shows experimental data for the deflection signal $u_d(t)$, and
the output $u_{y}(t)$ measured when the DSFM--DU is configured as just
discussed. The main graphs show spectra, and the insets time domain data. The
spectra have been acquired using a Spectrum Analyzer \cite{Audio} as well as
the ``noise'' function of a commercial Lock--In Amplifier \cite{Lockin}. The
lines through the data points correspond to a ``Lorentzian + offset'' fit
function $f(\nu)$ \cite{Ruido,ColcheroProc},
\begin{equation}
f(\nu)=\frac{e^2_{th}}{(1-(\nu/\nu_0)^2)^2+(\nu/(\nu_0Q))^2}+e^2_n=e^2_{th}|g(\nu)|^2+e^2_n\label{eq3}
\end{equation}
where $g(\nu)$ is the mechanical gain of the cantilever (see relation \eqref{eqA1}).
The corresponding parameters for the different fits are
specified in table \ref{Tab1}, and discussed in more detail below. The
scattered points around the horizontal lines show the error of the experimental
power spectrum to the corresponding fit, which show little tendency for both
spectra, therefore we conclude that the chosen fit functions describe well the
experimental data. Both spectra lead to essentially the same results, in
particular for the values of the resonance frequency $\nu_{0}$ and the quality
factor $Q$. Since the DSFM--DU internally amplifies the normal force signal, the
signal strength of the deflection signal $u_d(t)$ is smaller than that of the
DSFM--DU outputs $u_{x}(t)$ and $u_{y}(t)$, leading to different thermal noise
$e_{th}$ and electronic noise $e_{n}$ for the deflection spectrum $u_d(\nu)$ compared to
$u_{x}(\nu)$ and $u_{y}(\nu)$. We therefore conclude that the outputs $u_{x}(t)$
and $u_{y}(t)$ of the DSFM--DU essentially reproduce the thermal noise spectrum of the deflection signal and
that the DSFM--DU therefore ``sees'' (internally amplified) thermal noise
shifted to the frequency $\nu_{0}-\nu_{r}$. It should be therefore possible to
apply the well--known thermal calibration method based on the equipartition
theorem \cite{ColcheroProc,Hutter,Butt,Burnham,Gates,StarkThermal,SaderRoundRobin},
\begin{equation}
\frac{c}{2}\langle z_{th}^{2}(t)\rangle _{\mathrm{rms}}=\frac{1}{2}kT  \label{eq1}
\end{equation}
to the calibration of the oscillation amplitude.

We note that, as for the case of normal force calibration, relation \eqref{eq1}
has implicitly two unknowns: the force constant $c$ of the cantilever and a
sensitivity calibration $\beta$ (units: nm/V) to convert the measured signal $u(t)$
(Volts) into physical length units (Nanometers):
\begin{equation}
z(t)=\beta u(t)\label{eq0}
\end{equation}
In the present work, we will assume that the force constant is known or has been determined independently (see, for example \cite{Sader, SaderRoundRobin}), and will thus use \eqref{eq1} to calibrate the sensitivity $\beta$ of the DSFM--DU. We will present and discuss two procedures based on equation \eqref{eq1}: a simple one which only requires estimation of the RMS noise of the output of the DSFM--DU, and a second one taking into account the
full spectral response of the outputs $u_{x}(t)$ and $u_{y}(t)$.

\subsection{Calibration based on root mean square estimation}

The first calibration procedure is implemented experimentally by acquiring the outputs
$u_{x}(t)$ and $u_{y}(t)$ for a certain period of time with the laser on and
with the laser off. The noise power that is related to thermal noise is then,
\begin{equation}
\left(u_{th}\right)^{2}=\langle u_{\mathrm{on}}^{2}(t)\rangle_{\mathrm{rms}}-\langle u_{\mathrm{off}}^{2}(t)\rangle_{\mathrm{rms}}\label{eq2}
\end{equation}
The reason to acquire data with the laser off is to estimate and subtract
electronic noise that is not related to the thermal noise of the cantilever
motion. In particular for hard cantilevers, where thermal noise is quite small
(see below) the calibration may be severely wrong if this electronic noise is
not subtracted. Without substraction, it is implicitly assumed that all
measured noise (that is, also electronic noise) is thermal noise. The voltage
noise $\Delta u_{th}^{x}=\sqrt{\langle\left(u_{th}^{x}\right)^{2}\rangle}$ of
the output $u_{x}(t)$ is related to the ``phase'' noise while the noise $\Delta
u_{th}^{y}=\sqrt{\langle\left(u_{th}^{y}\right)^{2}\rangle}$ is related to the
oscillation amplitude. Relation \eqref{eq2} can of course also be applied to
the fluctuations measured in the deflection signal $u_d(t)$, then $\Delta
u_{th}^{d}=\sqrt{\langle\left(u_{th}^{d}\right)^{2}\rangle}$ is the deflection
noise (in Volts) of the cantilever motion. We recall that thermal noise is
distributed equally between the in--phase and out--of--phase components; then, if
the amplification of the outputs $u_{x}(t)$ and $u_{y}(t)$ as well as the
deflection $u_d(t)$ are the same (the latter is usually not the case, see below),
\begin{equation}
\Delta u_{th}^{x}=\Delta u_{th}^{y}=\Delta u_{th}^{d}/\sqrt{2}\label{eqraiz2}
\end{equation}
since the total deflection signal and the in--phase and out--out--phase componentes $x$ and $y$ are related by $u_d^2(t)=u_x^2(t)+u_y^2(t)$
and therefore $\beta^{x}=\beta^{y}$ (see, for example, \cite{Ruido}). The values $\Delta u_{th}^{x}$ and $\Delta u_{th}^{y}$ can be either computed directly from the time domain signals $u_{x}(t)$ and $u_{y}(t)$, or by adding (integrating) the noise spectrum in the frequency domain. Due to the different
normalizations found for the Fourier Transform and the Power Spectral Density in the literature as well as in software packets, it is usually more secure to use the first option. The insets presented in figure
\ref{Fig2} show the data used for this calibration procedure. The data from the
DSFM--DU is presented as scatter plot $(u_{x}(t),u_{y}(t))$; as it would be
visualized with an oscilloscope in $x-y$ mode. Data is shown for the laser on
and off. The diameter of this scatter plot is proportional to the fluctuations
in both directions. The data corresponding to the deflection signal $u_d(t)$ is
shown ``normally'' as a function of time, again data is shown for the laser on
and off. Calibration factors are computed directly as $\beta^{y}=z_{th}/\Delta
u_{th}^{y}/\sqrt{2}$ (see relation \eqref{eqraiz2} for the factor $\sqrt{2}$)
and $\beta^{d}=z_{th}/\Delta u_{th}^{d}$ with $z_{th}=\sqrt{kT/c}$ rms thermal noise movement (unit: nm, from equation \eqref{eq1}). The results
obtained with this first method are summarized in the left column of table
\ref{Tab1} (termed ``first procedure''). The calibration factors for the
cantilever deflection and for the (amplitude) output of the DSFM--DU have been
measured for a cantilever with $c=1.6$ N/m, as determined using Sader's method
\cite{Sader} from the resonance frequency and quality factor (see table
\ref{Tab1}).

\subsection{Calibration using the thermal noise spectrum}

The second calibration procedure is based on estimating the area under the
thermal noise spectrum using the values determined from the fit to a ``Lorentzian function + electronic noise function'', $f(\nu)$ from relation \eqref{eq3}.
In this context we recall that $\int_{0}^{\infty}d\nu \;\left\vert
g(\nu)\right\vert^{2}=~Q\ \nu _{0~}\pi /2$, therefore the
measured ``thermal noise power'' as determined from the fit parameters is
$e_{th}^{2}\ Q\ \nu_{0}\pi/2$, where $e^2_{th}$ is the strength of the thermal signal (units: V$^2$/Hz). As discussed previously, in the frequency domain
the DSFM--DU shifts the original deflection signal $z(\nu)$ from a signal
centered around the resonance frequency $\nu_{0}$ to signals $u_{x}(\nu)$ and
$u_{y}(\nu)$ centered at $\nu_{0}-\nu_{r}\approx bw/2$ where $\nu_r$ is the
reference frequency adjusted by the DSFM--DU. In addition, the
original signal $u_d(t)$ is amplified by the internal gains of the DSFM--DU. In
principle, there are two possible ways of processing the data obtained from the
DSFM--DU: the spectra $\left\vert u_{x}(\nu)\right\vert^{2}$ and $\left\vert
u_{y}(\nu)\right\vert^{2}$  may be fitted either directly to the function
$f(\nu)$ from relation \eqref{eq3}, resulting in a resonance peak near $bw/2$ (see figure \ref{Fig2}) --or the frequency shift
induced by the DSFM--DU may be ``undone'' by adding the reference frequency to
the frequency of the power spectra $u_{x}(\nu)$ and $u_{y}(\nu)$ in order to
fit these back--shifted power spectra $u_{x}(\nu+\nu _{r})$ and $u_{y}(\nu
+\nu_{r})$,
\begin{itemize}
\item Option 1: $u_{x}(t)$ raw data $\rightarrow$ FT and PSD
$\rightarrow\left\vert u_{x}(\nu )\right\vert^{2}\rightarrow
\rm{fit}\rightarrow\left\{e_{th},Q,\nu _{0},e_{n}\right\}$
\item Option 2: $u_{x}(t)$ raw data $\rightarrow${FT and PSD}$\rightarrow
\left\vert u_{x}(\nu )\right\vert^{2}\rightarrow$ freq.\
shift $\rightarrow|u_{x}(\nu +\nu _{r})|^{2}\rightarrow$ fit $\rightarrow\left\{e_{th},Q,\nu _{0},e_{n}\right\}$
\end{itemize}
The second option is the correct one, as relation \eqref{eq1} holds for the
``true''\ deflection signal with the thermal noise peak at the ``correct'' resonance
frequency $\nu_{0}$.

We therefore fit the frequency shifted spectrum $|u_{x}(\nu +\nu_{r})|^{2}$ to
the ``Lorentzian + offset'' function $f(\nu)$ from relation \eqref{eq3},
in order to obtain the ``strength'' $e_{th}$ of the Lorentz function, the
quality factor $Q$, the resonance frequency $\nu_{0}$ and the noise power $e_{n}$
that is not Lorentzian. For the experimental
spectra of the output $u_d(t)$, relation \eqref{eq1} is rewritten as
\begin{gather}
\begin{split}
\frac{1}{2}kT=\frac{c}{2}\langle z_{th}^{2}(t)\rangle_{\mathrm{rms}}=\frac{c}{2}%
&\int d\nu ~z_{th}^{2}\left\vert g(\nu)\right\vert^{2}=\frac{c}{2}\int d\nu ~\left(e_{th}^{d}~\beta^{d}\right)^{2}\left\vert
g(\nu)\right\vert^{2}=\\ 
&=\frac{c}{2}Q~\nu_{0}\frac{\pi}{2}
\left(e_{th}^{d}\beta^{d}\right)^{2}\Rightarrow \beta^{d}=\frac{1}{e_{th}^{d}}%
\sqrt{\frac{2kT}{\pi cQ\nu_{0}}}\label{eq4}
\end{split}
\end{gather}
where $\beta^{d}$ (unit: nm/Volts) is the factor that converts the deflection signal
$u_{d}(t) $ (in Volts) to the physical amplitude $z(t)$ (in nm):
$z(t)=\beta ^{d}u_{d}\left( t\right)$. Similarly, $\beta^{x}$ and $\beta^{y}$ convert the signals $u_{x}(t)$ and $u_{y}(t)$ into the in--phase and out--of--phase
components of cantilever deflection. For the outputs  $u_{x}(t)$ (and $u_{y}(t)$) relation \eqref{eq1} is
\begin{gather}
\begin{split}
&\frac{1}{2}kT =\frac{c}{2}\int d\nu
\left(\left(e_{th}^{x}~\beta ^{x}\right)^{2}+\left( e_{th}^{y}~\beta
^{y}\right)^{2}\right)\left\vert g(\nu +\nu_{r})\right\vert
^{2}=\\
&=2\frac{c}{2}Q\nu_{0}\frac{\pi}{2}\left(e_{th}^{y}~\beta
^{y}\right) ^{2}\Rightarrow \beta ^{y}=\frac{1}{\sqrt{2}e_{th}^{y}}\sqrt{%
\frac{2kT}{\pi cQ\nu _{0}}}  \label{eq5}
\end{split}
\end{gather}
The last relation for $\beta^{y}$ in \eqref{eq5} is obtained because in our case the amplification of the two
outputs $u_{x}(t)$ and $u_{y}(t)$ is the same; therefore, as discussed above,
$\beta^{x}=\beta^{y}$. The parameters $Q$, $\nu_{0}$ and $e_{th}$ are obtained
from the fits, $kT$ is the temperature of the cantilever, and $c=1.6$ N/m is
the force constant, determined by Sader's method \cite{Sader} from the $Q$
factor and the resonance frequency. The right columns of table \ref{Tab1}
(termed ``second procedure'') summarize the results obtained from this second
calibration procedure when applied to the deflection data $u_d(t)$ and to the out--of--phase
output $u_{y}(t)$ of the DSFM--DU. We will now comment the different
fields of this table. The Lock--In amplifier is used to measure the noise
density, and does not acquire time domain data, therefore the first procedure
cannot be applied, and the corresponding fields are empty. The Spectrum
Analyzer used allows to simultaneously measure the real--time data, as well as
the spectrum of this data, therefore the first and second procedure can be
applied. For this cantilever with a relatively large thermal noise (about 50
pm) the first and second calibration procedure give similar results, although
the second procedure seems to have less error (see discussion below). In order
to compare the results obtained from the deflection signal with that of the
output(s) of the DSFM--DU a normalized calibration factor $\beta_{N}$ has been
introduced that takes into account the internal gain of the DSFM--DU. This
normalized calibration factor should be the same for the deflection data and
the outputs of the DSFM--DU, which is indeed the case within the experimental
error of the measurements.

Figure \ref{Fig3} shows the spectra of the deflection sensor and the outputs of
the DSFM--DU for a hard cantilever ($c=67$ N/m as determined by Sader's
method \cite{Sader}), with a quite low thermal noise amplitude ($z_{th}=\sqrt{kT/c}\simeq
8$\ pm). In this case, data is shown for different gains of the DSFM--DU, and
the spectra have been acquired directly from the (calculated) power spectrum of
the time domain signals acquired by the SFM--control unit (more precisely: from
$u_{y}(t)$--images acquired with the tip far from the surface at maximum
adquisition speed, without scanning and no excitation applied to the driving
piezo of the cantilever). Note that for hard cantilevers, the resonance
frequency may be easily outside the acquisition bandwidth of the analog to
digital converters of the SFM--acquisition electronics (usually 16 bits or more
and thus rather slow) and therefore the corresponding thermal noise of
cantilever motion cannot be ``directly seen'' in the digitalized normal force
(deflection) signal $u_d(t)$. By contrast, the thermal noise peak is easily
brought into the bandwidth of the analog to digital converters when the outputs
$u_{x}(t)$ and $u_{y}(t)$ are used, since this noise is now at the much lower
difference frequency. For this hard cantilever, the very small thermal noise
signal is usually significantly smaller than the electronic noise or other non--thermal
fluctuations. In our setup, this is indeed the case for the direct
deflection signal as well as for the outputs of the DSFM--DU when its internal
gain is smaller than 10 (see the ``scatterplots'' in the inset of
figure \ref{Fig3}). Then, it is essential to ``normalize'' the measured signal by
subtracting the electronic noise (laser off) from the total noise (laser on) as
discussed above (see relation \eqref{eq2}). Still, for low gains this procedure
does not give satisfactory results; then only the second procedure method is
precise. Surprisingly the second procedure still works for gains as small as
$g=1$ and for the direct deflection signal. Note that for these low gains almost all noise is
electronic noise. The fit to the function $f(\nu )=e^2_{th}\left\vert
g\left(\nu\right) \right\vert^2 +e_{n}^2$ is thus a very effective way of
``filtering'' all non--thermal noise. For high gains the calibration factors rise
by an amount that is not compatible with the error of our measurements for both
procedures. We believe that this is not a problem of the calibration
procedures, instead we think that this is due to low--pass filtering of our
DSFM--DU that reduces the (nominal) signal strength and leads to a higher
calibration value at high frequency (more nanometers of deflection are
``needed'' for 1 V signal).

\section{Conclusion}

We have presented two methods for calibration of the oscillation amplitude in
DFSM, one based on simple calculation of the RMS value of the output of the
DSFM--DU $u_{y}(t)$, and the other one based on analysis of the corresponding
spectrum $u_{y}(\nu)$ and calculation of the parameters $Q$, $\nu_{0}$,
$e_{th}$ and $e_{n}$. From the results summarized in tables \ref{Tab1} and
\ref{Tab2} we conclude that both methods give consistent results, even though
the second method is considerably more precise and robust, in particular for
the case of hard cantilevers, where thermal noise has a much smaller amplitude
and the signal to noise ratio of thermal noise (which in this case is ``good''
signal) vs. other noise sources is much lower. The second method has several
advantages: first, it explicitly ``filters'' thermal noise from other noise
sources, since it will only take into account signal that has a Lorentzian
shape, signal not compatible with this shape is taken into account by the
constant factor $e_{n}$ and the corresponding signal power is rejected for the
calculation of the amplitude sensitivity.
Also, fitting of a spectrum with many data points in order to
obtain the four parameters $Q$, $\nu_{0}$, $e_{th}$ and $e_{n}$ results in
effective data averaging, and thus additional improvement of the estimation of
the thermal noise power. And finally, from a strictly theoretical point of
view, the first method is not correct because it is only an approximation valid
for high $Q$ factors. In fact, the first method only ``sees'' thermal noise in
the small bandwidth $1/\tau\ll\nu_{0}$ around the resonance peak, but not all
the noise under the Lorentz function, and in particular not the noise in the
low frequency ``tail'' (from DC to $\nu_{0}-bw/2$). This noise is ``filtered
away'' by the DSFM--DU and thus not taken into account; therefore the first
method underestimates noise, and overestimates the amplitude sensitivity (see
equation \eqref{eq1}). Since the relation of thermal noise in the resonance
peak to that in the low frequency ``tail'' is $Q:1$ (see for example
\cite{Ruido}, section 3) this error is negligible for experiments in air and
vacuum, but is expected to be significant for the low $Q$ factors encountered
in liquids. Finally, we note that determination of the parameters $e_{th}$, $Q$ and $\nu_0$ using the
outputs of the DSFM--DU involves significant improvement of signal because of the internal gains
of the DSFM--DU and the principle of Lock--in detection, which implies an important reduction
of bandwidth.

\section{Acknowledgments}

The authors acknowledge stimulating discussions with Ignacio Horcas, Arvin
Raman, Mariano Cuenca, Prashant Kulshreshtha and Luis Colchero. This work was
supported by the Spanish Ministry of Science and Technology as well as the European Union (FEDE founds) through the projects MAT2010-21267-C02-01 and CONSOLIDER Programme ``Force for Future'' (CSD2010-00024) as well as by the ``Comunidad Aut\'{o}noma de la Regi\'{o}n de Murcia'' through the project ``C\'{e}lulas
solares org\'{a}nicas: de la estructura molecular y nanom\'{e}trica a dispositivos
operativos macrosc\'{o}picos''. JFGM  and INC acknowledge doctoral grants to
the ``Ministerio de Eduaci\'{o}n'' (FPU programme and ``Fundaci\'{o}n S\'{e}neca'' of
the ``Comunidad Autónoma de la Regi\'{o}n de Murcia'', respectively).

\appendix

\section{Processing of thermal noise by a Dynamic Unit \label{AppendixA}}
%\long\def\symbolfootnote[#1]#2{\begingroup%
%\def\thefootnote{\fnsymbol{footnote}}\footnote[#1]{#2}\endgroup}

\begin{figure}[!ht]
\begin{center}
\includegraphics[width=18cm]{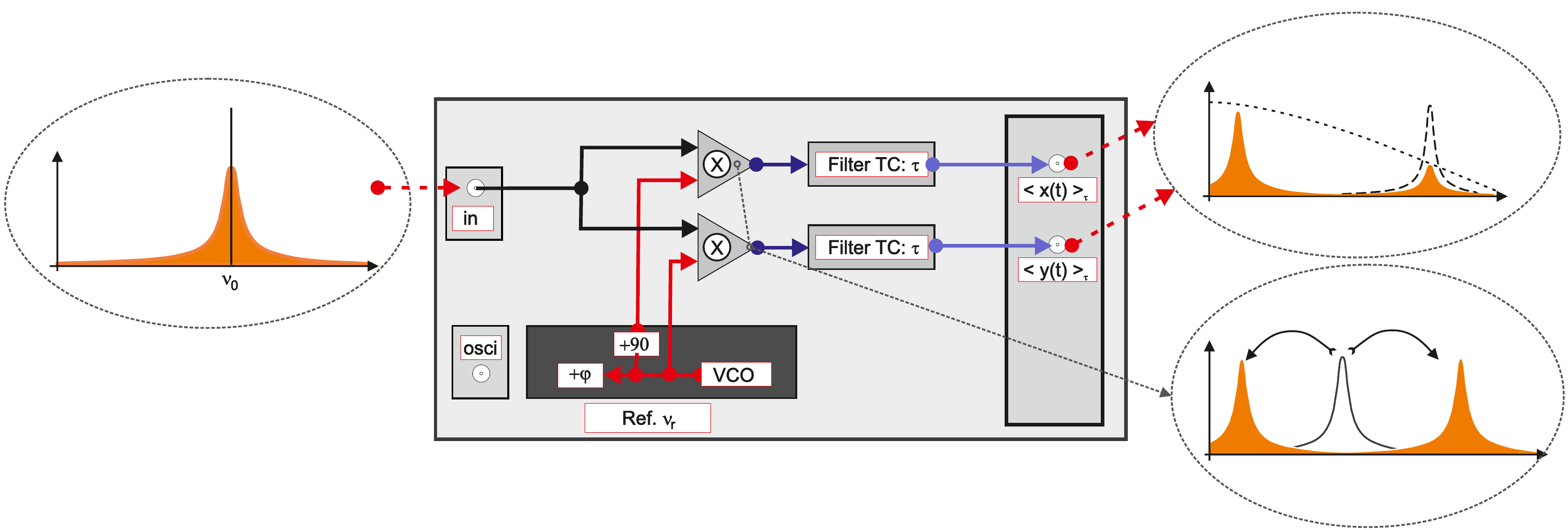}
\end{center}
\caption{Schematic description of a typical Lock--in type DSFM detection unit.
The signal to be analyzed by the DSFM detection is assumed to be centered
around some frequency $\nu_{0}$. It enters the detection unit at the input
``\textit{in}'', is amplified by some factor $g$ that can be generally selected
and usually high-pass filtered (for simplicity the corresponding components are
not shown) before it is multiplied with two reference signals in quadrature at
some frequency $\nu_{\rm{ref}}$. After this multiplication, the signal is
shifted to the frequencies $\nu_{0}-\nu_{\rm{ref}}$ and
$\nu_{0}+\nu_{\rm{ref}}$. The resulting signals are then low-pass filtered to
remove the higher frequency component ($\nu_{0}+\nu_{\rm{ref}}$), and brought
to the outputs $u_{x}(t)$ and $u_{y}(t)$.} \label{Fig1}
\end{figure}

\begin{figure}[!ht]
\begin{center}
\includegraphics[width=18cm]{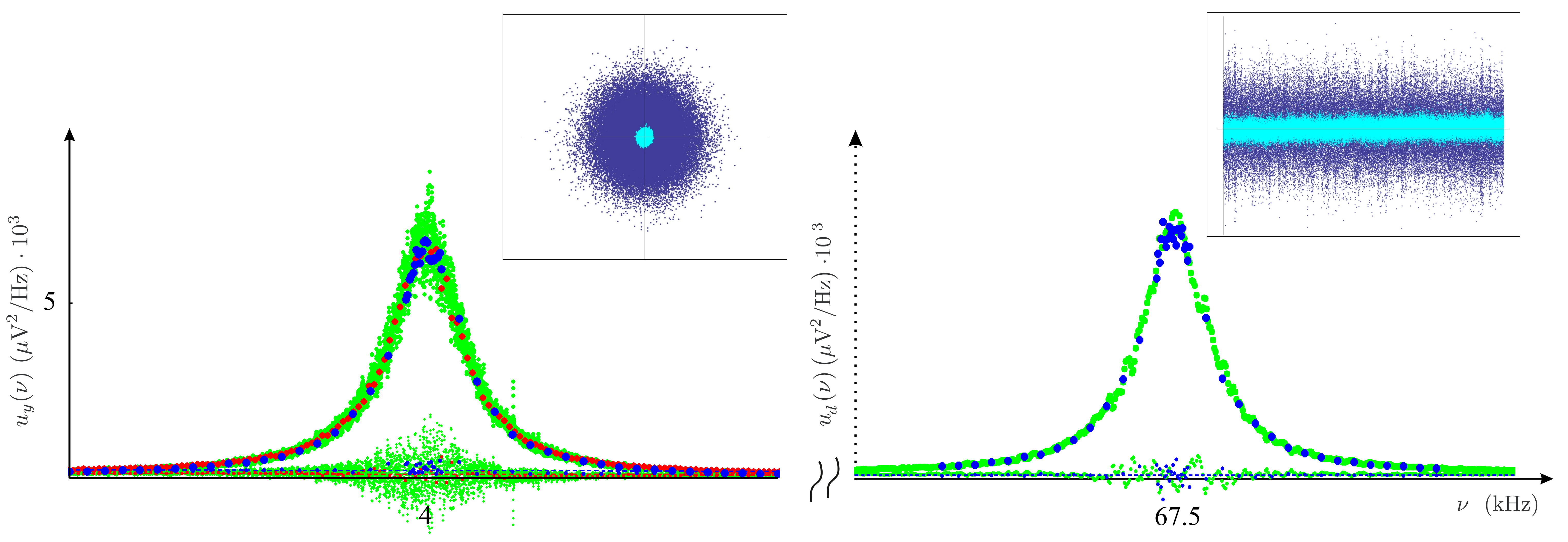}
\end{center}
\caption{Power spectra $u_{y}(\nu)$ (left curve) and $u_d(\nu )$ (right curve; this curve
is amplified to match the vertical range of the amplitude data) of the outputs
$u_{y}(t)$ of the DFSM-DU unit and the deflection signal $u_d(t)$. Data has been
acquired with an audio analyzer (green, small points), directly by the
AD--converters of our control electronics (red, larger points) and with a Lock--In
amplifier (blue, the largest points). Left insets: phase vs. amplitude,
$(u_{x}(t), u_{y}(t))$ data acquired with an audio analyzer. Right inset:
deflection vs. time signal, $u_d(t)$. For both insets: dark blue, signals
with the laser on; light blue, signals with the laser off.} \label{Fig2}
\end{figure}

\begin{figure}[!hb]
\begin{center}
\includegraphics[width=18cm]{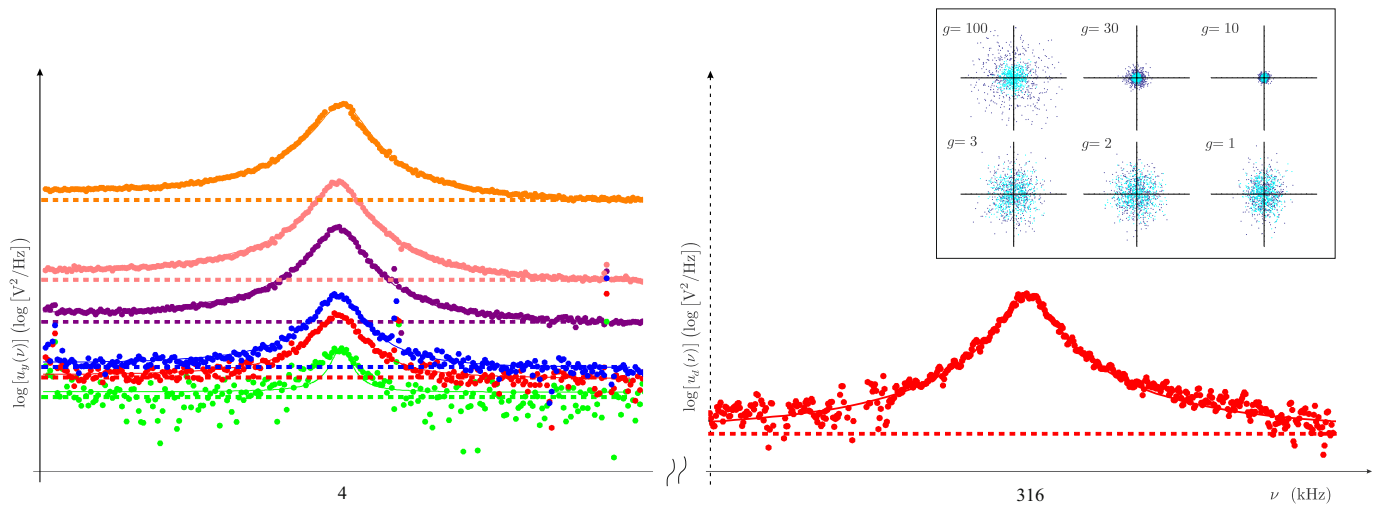}
\end{center}
\caption{Thermal noise data acquired for a hard cantilever ($c=67$ N/m). Left
side: power spectra of the outputs $u_{y}(t)$ of the DFSM--DU unit acquired
with different input gains. (a) Power spectra for gains $g =$ 1, 2, 3, 10, 30
and 100. Note that in this figure a logarithmic representation of the spectra
has been choosen in order to compress the vertical scale to fit all spectra
with very different gains. Lower right graph: Spectrum of the deflection vs.
time signal $u_d(t)$, measured with a digital oscilloscope. Inset: phase vs.
amplitude scatter plots $(u_{x}(t), u_{y}(t))$ from the data that was used to
compute the spectra $u_{y}(\nu)$ shown left; dark blue: signals with the laser
on, light blue: signals with the laser off. The upper three scatter plots of
the graph correspond to $(u_{x}(t), u_{y}(t))$ data for gains $g =$100, 30 and
10; the lower three scatter plots to data for gains $g =$ 3, 2 and 1.}
\label{Fig3}
\end{figure}

We recall that a typical electronics used for analyzing the dynamics of the
cantilever (see figure \ref{Fig1}), may be implemented using a Lock--in detection
scheme \cite{Albrecht,Durig,Ruido}: the input signal $u_d(t)$ to be analyzed is
multiplied by two reference signals in quadrature ($a_{r}\cos (2\pi \nu_{r}t)$
and $a_{r}\sin (2\pi \nu_{r}t)$) and then filtered with an appropriate time
constant $\tau$. The corresponding output signals of the DSFM--DU are the
in--phase (``phase'') and the out--of--phase (``amplitude'')
signals\footnote{Note that in \cite{Ruido} the nomenclature $\langle
x_{q}(t)\rangle_{\tau}$ and $\langle y_{q}(t)\rangle_{\tau }$ was used, that
is: $u_{x}(t)\equiv\langle x_{q}(t)\rangle_{\tau }$ and $u_{y}(t)\equiv\langle
y_{q}(t)\rangle_{\tau}$.}\ $ u_{x}(t)$ and $u_{y}(t)$. Generally in DSFM the
reference signal $a_{r}\cos (2\pi \nu_{r}t)$ is used to excite the cantilever at
resonance, then $u_{x}=0$; and changes of $u_{x}(t)$ are proportional to the
phase (and thus frequency) variation of the cantilever oscillation and
$u_{y}(t)$ is its oscillation amplitude. For FM--DSFM a feedback loop (Phase
Locked Loop, PLL) is used to track the resonance frequency of the tip--sample
system by changing the excitation (i.e., the reference) frequency, in order to
keep $u_{x}(t)=0$ (not shown in figure \ref{Fig1}). In a recent work \cite{Ruido}
we have analyzed in detail how such a DSFM--DU processes signals in the
presence of thermal noise. In particular, it was shown that the DSFM--DU
``frequency--shifts'' a signal $u_d(t)=\mathrm{Re}[a(\nu)e^{2\pi i\nu t}]$ with
\begin{equation}
a(\nu )=x(\nu )+iy(\nu)=\frac{a_{0}}{1-\left( \nu /\nu_{0}\right)
^{2}+i(\nu /\nu_{0})/Q}=a_{0}~g\left( \nu \right)\label{eqA1}
\end{equation}
at its input, to the sum and difference frequencies, where $g(\nu )$ is the complex
``mechanical gain''\ of the cantilever. The outputs ``phase'' and ``amplitude'' have frequency components
at $\nu_{\Sigma }=\nu_{0}+\nu_{r}$ and $\nu_{\Delta}=\nu_{0}-\nu_{r}$: $u_{x}(t)=\mathrm{\mathrm{Re}}\left[ M(t)\right]$
and $u_{y}(t)=\mathrm{\mathrm{Im}}\left[M(t)\right]$ with
\begin{equation}
M(t)=\frac{a(\nu )}{2}\left[\frac{1}{1+i2\pi \nu_{\Sigma}\tau}e^{-2\pi i(t~\nu _{\Sigma }+1/2)}+\frac{1}{1+i2\pi\nu_{\Delta}\tau}e^{+2\pi
i(t~\nu_{\Delta}+1/2)}\right] \label{eqA2}
\end{equation}
where, the matrix notation in \cite{Ruido}, appendix A, has been translated into the more compact complex
notation. When the input signal is not at a well defined frequency, but
distributed around a central frequency $\nu_{0}$, $u_d(t)=\int d\nu ~u_d(\nu
)e^{i2\pi \nu t}$, then the outputs of the DSFM--DU will be:
\begin{equation}
u_{x}(t)=\frac{1}{2}\mathrm{Re}\left[ D\left( t\right) \right] {\rm{\ and\ }}%
u_{y}(t)=\frac{1}{2}\mathrm{Im}\left[ D\left( t\right) \right]   \label{eqA3}
\end{equation}
with
\begin{equation}
D\left( t\right) =\int d\nu_{\Delta }d\left(\nu_{\mathrm{ref}}+\nu_{\Delta
}\right) e^{2\pi i~\nu_{\Delta }t}\frac{1}{1+i2\pi \nu_{\Delta}\tau}
\label{eqA4}
\end{equation}
where we have assumed that the term with the sum frequency $\nu_{\Sigma}$ (see
relation \eqref{eqA2}) can be neglected because the time constant $\tau $ of the
DSFM--DU is much larger than the time $1/\nu_{\Sigma}$ (see also
figure \ref{Fig1}). The DSFM--DU therefore shifts the spectrum of its input signal
$u_d(t)$ centered at $\nu_{0}$ to a frequency $\nu_{\Delta}=\nu_{0}-\nu_{r}$ of
the output signals $u_{x}(t)$ and $u_{y}(t)$. That is, if the power spectrum of
$u_d(t)$ has a maximum at $\nu_{0}$, the output signals $u_{x}(t)$ and $u_{y}(t)$
will have maxima at the much lower frequency $\nu_{\Delta}=\nu_{0}-\nu_{r}$, as stated in the main text of the present work.

\clearpage
\newpage

\section{Figures and Tables}

\clearpage
\newpage

\clearpage
\newpage

\renewcommand{\tabcolsep}{3.0pt} \renewcommand{\arraystretch}{1.8}
\begin{table}[bh]
{\scriptsize {\ }}
\par
\begin{center}
{\scriptsize
\begin{tabular}{cc|c|c|c|c||c|c|c|c|c|c|c}
\cline{3-12}
&  & \multicolumn{4}{|c||}{\textbf{First Procedure}} & \multicolumn{6}{|c|}{%
\textbf{Second Procedure}} &  \\ \cline{3-12}
&  & $u^{\rm{rms}}_{\rm{on}}$ & $u^{\rm{rms}}_{\rm{off}}$ & $u^{\rm{rms}}_{\rm{th}}$ & $\beta^{\rm{rms}}$ & $\nu$ & $Q$ & $e_{\rm{th}}$ & $e_n$ & $\beta^{\rm{PSD}}_{10}$ & $\beta^{\rm{PSD}}_{N}$ &  \\
\cline{2-12} & \multicolumn{1}{|c|}{Units $\rightarrow$} & mV & mV & mV & nm/V & Hz &  & V$/\sqrt{\rm{Hz}}\cdot10^{-6}$ & V$/\sqrt{\rm{Hz}}\cdot10^{-6}$ & nm/V & nm/V & \\
\cline{1-12}\multicolumn{1}{|c|}{\textbf{DSP}} & \multicolumn{1}{|c|}{$\langle u_y(t)\rangle$} & $5.38$ & $0.72$ & $5.33$ & $9.68$ & $67438.0(\pm14\rm{ ppm})$ & $138.6(\pm0.6\%)$ & $1.3(\pm0.5\%)$ & $30(\pm1.9\%)$ & $10.39(\pm0.8\%)$ & $36.7(\pm0.8\%)$ &  \\
\cline{1-12}\multicolumn{1}{|c|}{\multirow{2}{*}{\textbf{LI}}} & \multicolumn{1}{|c|}{$u_d(t)$} & $-$ & $-$ & $-$ & $-$ & $67440(\pm40\rm{ ppm})$ & $137(\pm2.4\%)$ & $0.35(\pm2.4\%)$ & $3(\pm66\%)$ & $-$ & $38.7(\pm3.6\%)$ &  \\
\cline{2-12}\multicolumn{1}{|c|}{} & \multicolumn{1}{|c|}{$\langle u_y(t)\rangle$} & $-$ & $-$ & $-$ & $-$ & $67440(\pm30\rm{ ppm})$ & $139(\pm1.6\%)$ & $1.3(\pm1.6\%) $ & $20(\pm13\%)$ & $10.35(\pm1.1\%)$ & $36.6(\pm2.4\%)$ &  \\
\cline{1-12}\multicolumn{1}{|c|}{\multirow{2}{*}{\textbf{AA}}} & \multicolumn{1}{|c|}{$u_d(t)$} & 1.77 & 0.39 & $1.72$ & $29.9$ & $67441(\pm160\rm{ ppm})$ & $140(\pm6.3\%)$ & $0.30(\pm5\%)$ & $5(\pm12\%)$ & $-$ & $39(\pm8\%)$ &  \\
\cline{2-12}\multicolumn{1}{|c|}{} & \multicolumn{1}{|c|}{$\langle u_y(t)\rangle$} & $5.40$ & $0.56$ & $5.37$ & $9.61$ & $67440(\pm5\rm{ ppm})$ & $137.8(\pm0.22\%)$ & $1.33(\pm0.2\%)$ & $20(\pm1.5\%)$ & $10.12(\pm0.3\%)$ & $35.8(\pm0.3\%)$ &  \\ \cline{1-12}
\end{tabular}
}
\end{center}
\par
\caption{Results obtained with the two calibration procedures applied to the
measured deflection and amplitude noise acquired from a cantilever with a
force constant $c=1.6$ N/m.}
\label{Tab1}
\end{table}

\renewcommand{\tabcolsep}{3.0pt} \renewcommand{\arraystretch}{1.8}
\begin{table}[bh]
\par
\begin{center}
{\scriptsize
\begin{tabular}{cc|c|c|c|c|c||c|c|c|c|c|c}
\cline{3-12}
&  & \multicolumn{5}{|c||}{\textbf{First Procedure}} & \multicolumn{5}{|c|}{\textbf{%
Second Procedure}} &  \\ \cline{3-12}
&  & $u^{\rm{rms}}_{\rm{on}}$ & $u^{\rm{rms}}_{\rm{off}}$ & $u^{\rm{rms}}_{\rm{th}}$ & $\beta^{\rm{rms}}$ & $\beta^{\rm{rms}}_N$ & $\nu$ & $Q$ & $e_{\rm{th}}$ & $\beta^{\rm{PSD}}_{G}$ & $\beta^{\rm{PSD}}_{N}$ &  \\
\cline{2-12}& \multicolumn{1}{|c|}{Units $\rightarrow$} & mV & mV & mV & nm/V & nm/V & Hz&  & V/$\sqrt{\rm{Hz}}$ $\cdot10^{-9}$ &  nm/V & nm/V &  \\
\cline{1-12}\multicolumn{1}{|c|}{\rm \bf{Osci}} & $\langle u_d(t)\rangle$ & $0.77$ & $0.68$& $0.36$ & -- & $20$ & $316066(\pm7\rm{ ppm})$ & $588(\pm1.3\%)$ & $8.6(\pm8\%)$ & -- & $53(\pm9\%)$ &  \\
\cline{1-12}\multicolumn{1}{|c|}{\multirow{4}{*}{\rm \bf{DSP}}} & \multicolumn{1}{|c|}{$\langle u_y(t)\rangle_{\rm{G}=1}$} & 0.314 & 0.299 & 0.095 & 80& 30& $316060(\pm120\rm{ ppm})$ & 580($\pm22\%$) & $3.2(\pm18\%)$ & $140(\pm29\%)$ & $51(\pm29\%)$ & \\ \cline{2-12}\multicolumn{1}{|c|}{} & \multicolumn{1}{|c|}{$\langle u_y(t)\rangle_{\rm{G}=2}$} & 0.33 & 0.29 & 0.16 & $50$ & $40$ & $316050(\pm73\rm{ ppm})$ & 570($\pm13\%$) & $6.5(\pm11\%)$ & $72(\pm18\%)$ & $50(\pm18\%)$&  \\
\cline{2-12}\multicolumn{1}{|c|}{} & \multicolumn{1}{|c|}{$\langle u_y(t)\rangle_{\rm{G}=3}$} & 0.36 & 0.30 & 0.21 & $40$ & $40$ & $316050(\pm59\rm{ ppm})$ & 570($\pm13\%$) & $9.4(\pm9\%)$ & $50(\pm14\%)$ & $51(\pm14\%)$&  \\
\cline{2-12}\multicolumn{1}{|c|}{} & \multicolumn{1}{|c|}{$\langle u_y(t)\rangle_{\rm{G}=10}$} & 0.73 & 0.38 & 0.63 & $13$ & $44$ & $316054(\pm6\rm{ ppm})$ & 570($\pm1.3\%$) & $33(\pm1\%)$ & $14.1(\pm1.7\%)$ & $50.1(\pm1.7\%)$ &  \\
\cline{2-12}\multicolumn{1}{|c|}{} & \multicolumn{1}{|c|}{$\langle u_y(t)\rangle_{\rm{G}=30}$} & 1.53 & 0.60 & 1.41 & $5.6$ & $48$ & $316053(\pm0.4\rm{ ppm})$ & 576($\pm0.67\%$) & $74.8(\pm0.56\%)$ & $6.21(\pm0.9\%)$ & $53.2(\pm0.9\%)$ &  \\ \cline{2-12}\multicolumn{1}{|c|}{} & \multicolumn{1}{|c|}{$\langle u_y(t)\rangle_{\rm{G}=100}$} & 6.24 & 2.02 & 5.91 & $1.33$ & $53$ & $316055(\pm0.3\rm{ ppm})$ & 563($\pm0.6\%$) & $316.2(\pm0.51\%)$ & $1.49(\pm0.8\%)$ & $52.9(\pm0.8\%)$ &  \\
\cline{1-12}
\end{tabular}
}
\end{center}
\par
\caption{Results obtained with the two calibration procedures for a hard\
cantilever (force constant $c=67$ N/m).}
\label{Tab2}
\end{table}

\clearpage
\newpage

\section*{References}

\bibliographystyle{unsrt}
\bibliography{PsdErrorAndTopo}

\end{document}